\documentclass[prb,twocolumn,showpacs]{revtex4}

\usepackage{graphicx}
\usepackage{dcolumn}
\usepackage{bm}
\usepackage{amsmath,amssymb}

\emergencystretch=\maxdimen
\begin{document}


\title{States induced in the single-particle spectrum by doping a Mott insulator}

\author{Masanori Kohno}
\affiliation{WPI Center for Materials Nanoarchitectonics, 
National Institute for Materials Science, Tsukuba 305-0044, Japan}

\date{\today}

\begin{abstract}
In strongly correlated electron systems, the emergence of states in the Mott gap in the single-particle spectrum following the doping of the Mott insulator 
is a remarkable feature that cannot be explained in a conventional rigid-band picture. 
Here, based on an analysis of the quantum numbers and the overlaps of relevant states, as well as through a demonstration using the ladder and bilayer $t$-$J$ models, 
it is shown that in a continuous Mott transition due to hole doping, the magnetically excited states of the Mott insulator generally emerge in the electron-addition spectrum 
with the dispersion relation shifted by the Fermi momentum in the momentum region where the lower Hubbard band is not completely filled. 
This implies that the dispersion relation of a free-electron-like mode in the electron-addition spectrum eventually transforms into essentially the momentum-shifted magnetic dispersion relation of the Mott insulator, 
while its spectral weight gradually disappears toward the Mott transition. 
This feature reflects the spin-charge separation of the Mott insulator.
\end{abstract}

\pacs{71.30.+h, 71.10.Fd, 74.72.Gh, 79.60.-i}

\maketitle
\section{Introduction} 
\label{sec:intro}
A significant feature in the Mott transition is the emergence of states in the Mott gap in the single-particle spectrum due to the doping of a Mott insulator. \cite{Eskes} 
Since such a feature does not appear when a noninteracting band insulator is doped, this phenomenon must be related to strong electronic correlations intrinsic in the Mott transition. 
However, although these states, which are called doping-induced states or in-gap states, have been observed in materials such as cuprate high-temperature (high-$T_{\rm c}$) superconductors and in various theoretical calculations, \cite{XraySWT_PRB,XraySWT_PRL,Takahashi_Bi2212,TaguchiPRL,ArmitageRMP,Eskes,Meinders,Eskes1D,EskesPRB,DagottoDOS,Dagotto_sumrule,DagottoAkwHub,DagottoRMP,ImadaRMP,Bulut,PreussQP,PreussPG,Stephan_nk,Ohta_ingapstates,MoreoQP,PhillipsRMP,PhillipsMottness,SakaiImadaPRL,SakaiImadaPRB,ImadaCofermionPRL,ImadaCofermionPRB,SakaiImadaRaman,Stanescu,KyungCDMFT,LiebschPRB,Tohyama2DtJ,Kohno1DHub,Kohno2DHub,KohnoSpin,Kohno2DtJ,Kohno2DHubNN,EderOhtaIPES,EderOhtaStripe,EderOhta2DHub,EderOhtaRigidband,Rice3legladder,DagottoAkwLadder,antiholon,ZemljicFiniteT,PrelovsekAw} 
their interpretations are controversial. 
\par
In particular, the question of whether the doping-induced states in the two-dimensional (2D) Hubbard and $t$-$J$ models in the parameter regime relevant to high-$T_{\rm c}$ cuprates near the Mott transition 
are essentially disconnected from the low-energy states by an energy gap or not has attracted considerable interest. 
This is because this issue is related not only to the nature of the Mott transition but also to the Fermi surface near the Mott transition (hole pockets or a free-electron-like Fermi surface)  \cite{DagottoRMP,PreussPG,Bulut,PreussQP,Stephan_nk,Ohta_ingapstates,SakaiImadaPRL,SakaiImadaPRB,ImadaCofermionPRL,ImadaCofermionPRB,Kohno2DHub,Kohno2DtJ,EderOhtaIPES,EderOhta2DHub,EderOhtaRigidband} and the anomalous features observed in high-$T_{\rm c}$ cuprates. \cite{DagottoRMP,ImadaRMP,ArmitageRMP} 
The presence of the energy gap has been suggested 
by numerical calculations \cite{SakaiImadaPRL,SakaiImadaPRB,SakaiImadaRaman,Tohyama2DtJ,PhillipsRMP,PhillipsMottness,EderOhtaIPES,EderOhta2DHub,EderOhtaRigidband} 
and interpreted in terms of various concepts, such as the binding between double occupancy and vacancy, \cite{SakaiImadaPRL,PhillipsRMP,PhillipsMottness,ImadaCofermionPRL,ImadaCofermionPRB} 
the effect of a charge 2$e$ boson, \cite{PhillipsRMP,PhillipsMottness} hybridization between the quasiparticle and cofermion, \cite{ImadaCofermionPRL,ImadaCofermionPRB} 
and a spin-polaron shakeoff. \cite{EderOhta2DHub,EderOhtaIPES} 
The absence of the energy gap has also been suggested by numerical calculations, \cite{Bulut,PreussQP,DagottoAkwHub,Stephan_nk,Ohta_ingapstates,Kohno2DHub,KohnoSpin,Kohno2DtJ} 
where the doping-induced states have been interpreted as a quasiparticle coupled with antiferromagnetic correlations \cite{PreussQP} 
and as essentially the magnetic excitation of the Mott insulator which emerges in the single-particle spectrum because charge character 
is added by doping. \cite{Kohno2DHub,KohnoSpin,Kohno2DtJ} 
\par
In this paper, we address this issue from a general perspective, by investigating how the dispersion relation of the doping-induced states should generally behave in the small-doping limit of a continuous Mott transition, 
based on an analysis of the quantum numbers and the overlaps of relevant states. 
The validity of the argument is demonstrated using the two-leg ladder and bilayer $t$-$J$ models. An intuitive real-space picture of the doping-induced states is also presented. 
\par
We consider zero-temperature properties of repulsively interacting systems where the ground states are Mott insulators at half-filling 
and metals in the small-doping limit. 
The numbers of sites, unit cells, electrons, and doped holes are denoted by $N_{\rm s}$, $N_{\rm u}$, $N_{\rm e}$, and $N_{\rm h}(=N_{\rm s}-N_{\rm e})$, respectively. 
The doping concentration is denoted by $\delta(=N_{\rm h}/N_{\rm s})$. 
\section{Quantum numbers} 
We study the single-particle spectral function defined as 
$$
A({\bm k},\omega)=\left\{
\begin{array}{lll}
\frac{1}{2}\sum_{m,\sigma}|\langle m|c^{\dagger}_{{\bm k},\sigma}|{\rm GS}\rangle_h|^2\delta(\omega-\varepsilon_m)&\mbox{for}&\omega>0,\\
\frac{1}{2}\sum_{m,\sigma}|\langle m|c_{{\bm k},\sigma}|{\rm GS}\rangle_h|^2\delta(\omega+\varepsilon_m)&\mbox{for}&\omega<0,
\end{array}\right.
$$
where $|{\rm GS}\rangle_h$ and $\varepsilon_m$ denote the ground state at $N_{\rm h}=h$ and the excitation energy of the eigenstate $|m\rangle$ from $|{\rm GS}\rangle_h$, respectively. 
Here, $c^{\dagger}_{{\bm k},\sigma}$ denotes the creation operator of an electron with momentum ${\bm k}$ and magnetization $\sigma$. 
\par
If $|{\rm GS}\rangle_h$ has spin $s$, magnetization $s_z$, and momentum ${\bm k}_{\rm F}$, $\langle m|c^{\dagger}_{{\bm k},\sigma}|{\rm GS}\rangle_h$ can be nonzero 
only for $|m\rangle$ with $h-1$ holes, spin $|s\pm1/2|$, magnetization $s_z+\sigma$, and momentum ${\bm k}+{\bm k}_{\rm F}$. \cite{Nozieres} 
In addition, if the ground-state energy (including the chemical-potential term) at $N_{\rm h}=h$ is lower than that of $N_{\rm h}=h-1$ by $\epsilon$, 
$N_{\rm e}$-conserving excited states with excitation energy $\omega-\epsilon$ from $|{\rm GS}\rangle_{h-1}$ can contribute to $A({\bm k},\omega)$ at $N_{\rm h}=h$, 
where $\epsilon\rightarrow +0$ for $N_{\rm s}\rightarrow\infty$ in a metallic phase (including $h=1$). 
In interacting systems, it is generally expected that low-energy eigenstates (particularly those of dominant excitations) 
having the same quantum numbers as those of components of $c^{\dagger}_{{\bm k},\sigma}|{\rm GS}\rangle_h$ 
have nonzero overlap with $c^{\dagger}_{{\bm k},\sigma}|{\rm GS}\rangle_h$ for ${\bm k}$ where the lower Hubbard band (LHB) is not completely filled. 
\par
Thus, if $|{\rm GS}\rangle_1$ has spin 1/2 and momentum ${\bm k}_{\rm F}$, $|{\rm GS}\rangle_0$ with spin 0 and momentum ${\bm 0}$ 
should contribute to $A({\bm k},\omega)$ at ${\bm k}=-{\bm k}_{\rm F}$ and $\omega=\epsilon$ for the top of the hole-like band [LHB at half-filling ($N_{\rm h}=0$)] in the small-doping limit, 
which causes hole-pocket-like behavior as in a doped band insulator. \cite{EderOhtaIPES} 
In addition, the magnetically excited states with spin 1 at half-filling \cite{EderOhtaIPES} having the dispersion relation $\omega=f({\bm k})$ should emerge following doping 
along $\omega=f({\bm k}+{\bm k}_{\rm F})+\epsilon$ in $A({\bm k},\omega)$ for ${\bm k}$ where the LHB is not completely filled. 
\par
According to this argument, if the magnetic excitation of the Mott insulator is gapless at momenta ${\bm 0}$ and ${\bm Q}$, 
the doping-induced states in the small-doping limit should also be gapless at $-{\bm k}_{\rm F}$ and ${\bm Q}-{\bm k}_{\rm F}$ ($\pm{\bm k}_{\rm F}$ if ${\bm Q}=2{\bm k}_{\rm F}$) and 
exhibit the momentum-shifted magnetic dispersion relation 
in the momentum region where the LHB is not completely filled (primarily outside the Fermi surface). 
This behavior essentially agrees with the results for the one-dimensional (1D) and 2D Hubbard models in Refs. \onlinecite{Kohno1DHub,Kohno2DHub,KohnoSpin} and 
those for the 2D $t$-$J$ model in Ref. \onlinecite{Kohno2DtJ} 
and suggests the coexistence of hole-like behavior and electron-like behavior for $\omega\gtrsim 0$. 
\par
Although it has been pointed out that the spin-0 and spin-1 states at half-filling can appear in the electron-addition spectrum at $N_{\rm h}=1$, 
the relationship of the momenta (dispersion relations) has not been recognized in Ref. \onlinecite{EderOhtaIPES}; 
the doping-induced states in the 2D $t$-$J$ and Hubbard models have been considered to be disconnected by an energy gap 
from the low-energy states relevant to hole-like Fermi surfaces and have been interpreted as spin-polaron shakeoff bands in Refs. \onlinecite{EderOhtaIPES,EderOhta2DHub}. 
However, according to the above argument, it is natural to interpret the doping-induced states in the small-doping limit 
as essentially the states of the spin-wave mode of the Mott insulator 
and to consider that they exhibit the gapless spin-wave dispersion relation \cite{AndersonSW,Manousakis2dHeis,Hirsch2dHub} shifted by ${\bm k}_{\rm F}$ in the 2D $t$-$J$ and Hubbard models. 
\section{Overlap} 
To strengthen the above argument, 
we consider the $t$-$J$ model defined by the following Hamiltonian: 
$$
{\cal H}=\sum_{i\ne j,\sigma}t_{i,j}{\tilde c}^{\dagger}_{i,\sigma}{\tilde c}_{j,\sigma}
+\sum_{i\ne j}J_{i,j}({\bm S}_i\cdot{\bm S}_j-\frac{1}{4}n_in_j)-\mu\sum_in_i,
$$
where ${\tilde c}_{i,\sigma}=c_{i,\sigma}(1-n_{i,-\sigma})$ and $n_i=\sum_{\sigma}n_{i,\sigma}$ 
for the annihilation operator $c_{i,\sigma}$ and number operator $n_{i,\sigma}$ of an electron with magnetization $\sigma$ at site $i$. 
Here, ${\bm S}_i$ denotes the spin operator at site $i$. 
At each site, a vacant state $|0\rangle$ or a singly-occupied magnetization-$\sigma$ state $|\sigma\rangle$ is allowed, but double occupancy is forbidden. 
Here, we assume that $|{\rm GS}\rangle_0$ has spin 0 and momentum ${\bm 0}$ and that $|{\rm GS}\rangle_1$ has spin 1/2, magnetization $\varsigma$, and momentum ${\bm k}_{\rm F}$. 
For instance, $|{\rm GS}\rangle_0$ on a bipartite lattice with the same number of sites in the two sublattices has spin 0 in a finite-size system with $J_{i,j}>0$ for neighboring sites between the two sublattices and $J_{i,j}=0$ otherwise. \cite{LiebMattis} 
\par
Because ${\tilde c}^{\dagger}_{i,\sigma}|\alpha\rangle^i$ can be expressed as $S^{+,\sigma}_i{\tilde c}^{\dagger}_{i,-\sigma}|\alpha\rangle^i$ and $2S^{z,\sigma}_i{\tilde c}^{\dagger}_{i,\sigma}|\alpha\rangle^i$ using 
$$
\begin{array}{ll}
S_i^{+,\sigma}={\tilde c}_{i,\sigma}^{\dagger}{\tilde c}_{i,-\sigma},&S_i^{z,\sigma}=(n_{i,\sigma}-n_{i,-\sigma})/2,
\end{array}
$$
where $|\alpha\rangle^i$ denotes the state at site $i$ 
(${\tilde c}^{\dagger}_{i,\sigma}|\alpha\rangle^i=|\sigma\rangle^i$ for $|\alpha\rangle^i=|0\rangle^i$, and 0 otherwise), 
\begin{equation}
\begin{array}{ccl}
_0\langle{\rm GS}|{\tilde c}^{\dagger}_{-{\bm k}_{\rm F},-\varsigma}|{\rm GS}\rangle_1
&=&\frac{1}{\sqrt{N_{\rm s}}}\sum_i{}_0\langle{\rm GS}|S_{-{\bm k}_{\rm F}-{\bm k}_i}^{+,-\varsigma}{\tilde c}^{\dagger}_{{\bm k}_i, \varsigma}|{\rm GS}\rangle_1\\
&=&\frac{2}{\sqrt{N_{\rm s}}}\sum_i{}_0\langle{\rm GS}|S_{-{\bm k}_{\rm F}-{\bm k}_i}^{z,-\varsigma}{\tilde c}^{\dagger}_{{\bm k}_i,-\varsigma}|{\rm GS}\rangle_1, 
\end{array}
\label{eq:overlap0}
\end{equation}
where 
$$
\begin{array}{ll}
{\tilde c}^{\dagger}_{{\bm k},\sigma}=\frac{1}{\sqrt{N_{\rm s}}}\sum_{j}{\rm e}^{\i{\bm k}\cdot{\bm r}_j}{\tilde c}^{\dagger}_{j,\sigma},
&S_{\bm k}^{+(z),\sigma}=\frac{1}{\sqrt{N_{\rm s}}}\sum_{j}{\rm e}^{\i{\bm k}\cdot{\bm r}_j}S^{+(z),\sigma}_j
\end{array}
$$
using the site $j$ coordinate ${\bm r}_j$. 
Equation (1) means that the overlaps between the electron-addition states at $N_{\rm h}=1$ (${\tilde c}^{\dagger}_{{\bm k},\varsigma}|{\rm GS}\rangle_1$ 
and ${\tilde c}^{\dagger}_{{\bm k},-\varsigma}|{\rm GS}\rangle_1$) 
and the single-spin excited states at half-filling ($S_{{\bm k}+{\bm k}_{\rm F}}^{+,\varsigma}|{\rm GS}\rangle_0$ and $S_{{\bm k}+{\bm k}_{\rm F}}^{z,-\varsigma}|{\rm GS}\rangle_0$) 
are related to the overlap between the electron-removal state for the momentum at the top of the LHB at half-filling (${\tilde c}_{-{\bm k}_{\rm F},-\varsigma}|{\rm GS}\rangle_0$) and $|{\rm GS}\rangle_1$. 
\par
From Eq. (\ref{eq:overlap0}), if ${}_0\langle{\rm GS}|{\tilde c}^{\dagger}_{-{\bm k}_{\rm F},-\varsigma}|{\rm GS}\rangle_1$ is $O(1)$, 
${}_0\langle{\rm GS}|S_{-{\bm k}_{\rm F}-{\bm k}}^{+,-\varsigma}{\tilde c}^{\dagger}_{{\bm k}, \varsigma}|{\rm GS}\rangle_1$ and 
${}_0\langle{\rm GS}|S_{-{\bm k}_{\rm F}-{\bm k}}^{z,-\varsigma}{\tilde c}^{\dagger}_{{\bm k},-\varsigma}|{\rm GS}\rangle_1$ are expected to be $O(1/{\sqrt N_{\rm s}})$ at $O(N_{\rm s})$ ${\bm k}$-points. 
Then, for these ${\bm k}$-points, the normalized electron-addition states should normally have $O(1)$ overlap with the normalized single-spin excited states 
because $\frac{1}{N_{\rm s}}\sum_i{}_0\langle{\rm GS}|{\bm S}_{-{\bm k}_i}\cdot{\bm S}_{{\bm k}_i}|{\rm GS}\rangle_0=3/4$ and 
$\frac{1}{N_{\rm s}}\sum_i{}_1\langle{\rm GS}|{\tilde c}_{{\bm k}_i,\sigma}{\tilde c}^{\dagger}_{{\bm k}_i,\sigma}|{\rm GS}\rangle_1=1/{N_{\rm s}}$. \cite{Stephan_nk,Dagotto_sumrule} 
\par
In addition, because all sites are singly occupied at half-filling, 
${\tilde c}^{\dagger}_{i,\sigma}{\tilde c}_{j,\sigma^{\prime}}|{\rm GS}\rangle_0=\delta_{i,j}{\tilde c}^{\dagger}_{i,\sigma}{\tilde c}_{i,\sigma^{\prime}}|{\rm GS}\rangle_0$ and 
$\frac{1}{N_{\rm s}}\sum_{j}{\rm e}^{\i({\bm k}-{\bm k}_{\rm F})\cdot{\bm r}_j}n_j|{\rm GS}\rangle_0=\delta_{{\bm k},{\bm k}_{\rm F}}|{\rm GS}\rangle_0$. 
Hence, 
$$
\begin{array}{lll}
{}_0\langle{\rm GS}|S_{-{\bm k}_{\rm F}-{\bm k}}^{+,-\varsigma}{\tilde c}^{\dagger}_{{\bm k}, \varsigma}|{\rm GS}\rangle_1&=&\sqrt{N_{\rm s}}{}_0\langle{\rm GS}|{\tilde c}^{\dagger}_{-{\bm k}_{\rm F},-\varsigma}{\tilde c}_{{\bm k},\varsigma}{\tilde c}^{\dagger}_{{\bm k},\varsigma}|{\rm GS}\rangle_1, \nonumber\\
{}_0\langle{\rm GS}|S_{-{\bm k}_{\rm F}-{\bm k}}^{z,-\varsigma}{\tilde c}^{\dagger}_{{\bm k},-\varsigma}|{\rm GS}\rangle_1&=&\sqrt{N_{\rm s}}{}_0\langle{\rm GS}|{\tilde c}^{\dagger}_{-{\bm k}_{\rm F},-\varsigma}{\tilde c}_{{\bm k},-\varsigma}{\tilde c}^{\dagger}_{{\bm k},-\varsigma}|{\rm GS}\rangle_1\nonumber\\
&-&\delta_{{\bm k},-{\bm k}_{\rm F}}\sqrt{N_{\rm s}}{}_0\langle{\rm GS}|{\tilde c}^{\dagger}_{-{\bm k}_{\rm F},-\varsigma}|{\rm GS}\rangle_1/2.
\end{array}
$$
Thus, the overlap can be roughly estimated in a decoupling approximation as 
\begin{equation}
\begin{array}{ccl}
{}_0\langle{\rm GS}|S_{-{\bm k}_{\rm F}-{\bm k}}^{+,-\varsigma}{\tilde c}^{\dagger}_{{\bm k}, \varsigma}|{\rm GS}\rangle_1&\approx&\sqrt{N_{\rm s}}\xi{}_1\langle{\rm GS}|{\tilde c}_{{\bm k},\varsigma}{\tilde c}^{\dagger}_{{\bm k},\varsigma}|{\rm GS}\rangle_1,\\
{}_0\langle{\rm GS}|S_{-{\bm k}_{\rm F}-{\bm k}}^{z,-\varsigma}{\tilde c}^{\dagger}_{{\bm k},-\varsigma}|{\rm GS}\rangle_1&\approx&\sqrt{N_{\rm s}}\xi{}_1\langle{\rm GS}|{\tilde c}_{{\bm k},-\varsigma}{\tilde c}^{\dagger}_{{\bm k},-\varsigma}|{\rm GS}\rangle_1\\
&&-\delta_{{\bm k},-{\bm k}_{\rm F}}\sqrt{N_{\rm s}}\xi/2,
\end{array}
\label{eq:overlap2}
\end{equation}
where $\xi={}_0\langle{\rm GS}|{\tilde c}^{\dagger}_{-{\bm k}_{\rm F},-\varsigma}|{\rm GS}\rangle_1$. At ${\bm k}=-{\bm k}_{\rm F}$, 
$$
{}_0\langle{\rm GS}|S_{-{\bm k}_{\rm F}-{\bm k}}^{+,-\varsigma}{\tilde c}^{\dagger}_{{\bm k}, \varsigma}|{\rm GS}\rangle_1
={}_0\langle{\rm GS}|S_{-{\bm k}_{\rm F}-{\bm k}}^{z,-\varsigma}{\tilde c}^{\dagger}_{{\bm k},-\varsigma}|{\rm GS}\rangle_1=0
$$
because $|{\rm GS}\rangle_0$ is assumed to be a spin-0 state. 
\par
The above results imply that the single-spin excited states of the Mott insulator contribute significantly to the doping-induced states in the small-doping limit 
for ${\bm k}$ where the LHB is not completely filled (${\tilde c}^{\dagger}_{{\bm k},\sigma}|{\rm GS}\rangle_1\ne 0$) and 
for ${\bm k}\ne -{\bm k}_{\rm F}$ if ${}_0\langle{\rm GS}|{\tilde c}^{\dagger}_{-{\bm k}_{\rm F},-\varsigma}|{\rm GS}\rangle_1$ is $O(1)$. 
In 1D systems, even if ${}_0\langle{\rm GS}|{\tilde c}^{\dagger}_{-{\bm k}_{\rm F},-\varsigma}|{\rm GS}\rangle_1\rightarrow 0$ for $N_{\rm s}\rightarrow \infty$, \cite{UinfSorella,Z1dSorella} 
exact-solution analyses and numerical calculations strongly suggest that the mode of the doping-induced states in the small-doping limit 
primarily reflects the dominant part of the magnetic excitation. \cite{Kohno1DHub,UinfPencPRB} 
\section{Weakly coupled two-site clusters} 
\label{sec:2siteClusters}
To demonstrate the validity of this argument, we consider the two-leg ladder and square-lattice bilayer $t$-$J$ models 
for $t_{\perp}\gg t_{\parallel}$, $J_{\perp}\gg J_{\parallel}$, $t_{\perp}\gg J_{\perp}$, and $t_{\parallel}\gg J_{\parallel}>0$ 
($J_{\perp}/t_{\parallel}$ is not too small for the ground state to have spin 0 or 1/2 [\onlinecite{KohnoUinf,KrivnovUinf}]). 
Here, $t_{\perp}$ and $J_{\perp}$ denote $-t_{i,j}$ and $2J_{i,j}$ for neighboring sites between chains (planes), respectively, 
and $t_{\parallel}$ and $J_{\parallel}$ denote $-t_{i,j}$ and $2J_{i,j}$ for neighboring sites in chains (planes), respectively, in the ladder (bilayer) model. 
The other $t_{i,j}$ and $J_{i,j}$ are set to zero. 
Although some aspects of the doping-induced states in similar models have been discussed, \cite{EderOhtaStripe,Rice3legladder,DagottoAkwLadder} 
this paper clarifies the nature of the doping-induced states through explicit calculations of the spectral weights, dispersion relation, and overlaps with magnetically excited states, 
using the effective eigenstates up to $O(t_{\parallel})$ and $O(J_{\parallel})$. 
\subsection{One-hole doping} 
The ground state at half-filling (spin 0 and momentum ${\bm 0}$), the low-energy spin-1 eigenstate at half-filling, and the low-energy eigenstate at $N_{\rm h}=1$ are expressed, respectively, as 
$$
\begin{array}{lll}
|{\rm GS}\rangle_0&=&|{\rm S}\rangle^1\otimes\cdots\otimes|{\rm S}\rangle^{N_{\rm u}}, \\
|{\rm T}\rangle_0^{{\bm k}_{\parallel},s_z}&=&\frac{1}{\sqrt{N_{\rm u}}}\sum_j{\rm e}^{\i{\bm k}_{\parallel}\cdot{\bm r}_{\parallel j}}|{\rm T}_{s_z}\rangle^j\otimes_{l\ne j}|{\rm S}\rangle^l, \\
|{\rm B}\rangle_1^{{\bm k}_{\parallel},s_z}&=&\frac{1}{\sqrt{N_{\rm u}}}\sum_j{\rm e}^{\i{\bm k}_{\parallel}\cdot{\bm r}_{\parallel j}}|{\rm B}_{s_z}\rangle^j\otimes_{l\ne j}|{\rm S}\rangle^l, 
\end{array} 
$$
where $|{\rm S}\rangle^j$, $|{\rm T}_{s_z}\rangle^j$, and $|{\rm B}_{s_z}\rangle^j$ represent the two-electron spin-0 state, spin-1 state, and one-electron bonding state, respectively, 
at the $j$-th two-site unit cell with coordinate ${\bm r}_{\parallel j}$ for $t_{\parallel}=J_{\parallel}=0$. 
Here, $s_z$ and ${\bm k}_{\parallel}$ denote the magnetization and inter-unit-cell momentum, respectively. 
The magnetic dispersion relation at half-filling is obtained by calculating the excitation energy of $|{\rm T}\rangle_0^{{\bm k}_{\parallel},s_z}$ from $|{\rm GS}\rangle_0$: 
$$
\Delta E({\bm k}_{\parallel})=\left\{
\begin{array}{ll}
J_{\parallel}\cos k_{\parallel}+J_{\perp}&\mbox{for 1D systems,}\\
J_{\parallel}(\cos k_{\parallel x}+\cos k_{\parallel y})+J_{\perp}&\mbox{for 2D systems.}
\end{array}\right.
$$
The ground state at $N_{\rm h}=1$ with magnetization $\varsigma$ ($|{\rm GS}\rangle_1$) is $|{\rm B}\rangle_1^{{\bm \pi},\varsigma}$ if ${\bm k}_{\parallel}$ can be set to ${\bm \pi}$, where ${\bm \pi}$ indicates $\pi$ and $(\pi,\pi)$ for 1D and 2D systems, respectively. 
For the small-doping limit, the value of $\mu$ is set such that ${}_1\langle{\rm GS}|{\cal H}|{\rm GS}\rangle_1={}_0\langle{\rm GS}|{\cal H}|{\rm GS}\rangle_0-\epsilon$ with $\epsilon\rightarrow +0$. 
\par
We calculate the spectral function at $N_{\rm h}=1$. By adding an electron to the ground state at $N_{\rm h}=1$, spin-0 and 1 states are obtained \cite{EderOhtaStripe} as follows: 
$|{\rm GS}\rangle_0/\sqrt{2}(={\tilde c}^{\dagger}_{-{\bm k}_{\rm F},-\varsigma}|{\rm GS}\rangle_1$) for spin $0$, and 
$|{\rm T}\rangle_0^{{\bm k}_{\parallel}+{\bm \pi},2\varsigma}/\sqrt{N_{\rm u}}(={\tilde c}^{\dagger}_{({\bm k}_{\parallel},\pi),\varsigma}|{\rm GS}\rangle_1$) and $|{\rm T}\rangle_0^{{\bm k}_{\parallel}+{\bm \pi},0}/\sqrt{2N_{\rm u}}(={\tilde c}^{\dagger}_{({\bm k}_{\parallel},\pi),-\varsigma}|{\rm GS}\rangle_1$) for spin $1$. 
Here, $({\bm k}_{\parallel},k_{\perp})$ indicates the momentum with inter- and intra-unit-cell momenta ${\bm k}_{\parallel}$ and  $k_{\perp}$, respectively, and ${\bm k}_{\rm F}=({\bm \pi},0)$. 
Thus, there are spectral weights at ${\bm k}_{\parallel}=-{\bm \pi}$ and $\omega=\epsilon$ in $A_{\rm b}({\bm k}_{\parallel},\omega)$ 
[$\frac{1}{2}\sum_{\sigma}|{}_0\langle{\rm GS}|{\tilde c}^{\dagger}_{{\bm k},\sigma}|{\rm GS}\rangle_1|^2=\delta_{{\bm k},-{\bm k}_{\rm F}}/4$] 
and essentially along the dispersion relation of the magnetic excitation shifted by $\bm \pi$, 
$\omega=\Delta E({\bm k}_{\parallel}+{\bm \pi})+\epsilon$, in $A_{\rm a}({\bm k}_{\parallel},\omega)$ 
[$\frac{1}{2}\sum_{s_z,\sigma}|{}_0^{{\bm k}_{\parallel}+{\bm \pi},s_z}\langle{\rm T}|{\tilde c}^{\dagger}_{({\bm k}_{\parallel},\pi),\sigma}|{\rm GS}\rangle_1|^2=3/(4N_{\rm u})$], 
where $A_{\rm b}({\bm k}_{\parallel},\omega)$ and $A_{\rm a}({\bm k}_{\parallel},\omega)$ denote the spectral functions for ${\tilde c}^{(\dagger)}_{({\bm k}_{\parallel},0),\sigma}$ and 
${\tilde c}^{(\dagger)}_{({\bm k}_{\parallel},\pi),\sigma}$, respectively. 
The shift of ${\bm \pi}$ is due to the momentum difference between $|{\rm GS}\rangle_1$ and $|{\rm GS}\rangle_0$. 
\par
The spectral weight for $\omega>0$ averaged over the momentum is $1/N_{\rm s}(=\delta)$, 
which is consistent with the spectral-weight sum rule. \cite{Stephan_nk,Dagotto_sumrule} 
One quarter of this spectral weight ($\delta/4$) is from the ground state at half-filling, which contributes to the mode causing hole-pocket-like behavior, 
whereas three quarters ($3\delta/4$) is from the magnetically excited states ($|{\rm T}\rangle_0^{{\bm k}_{\parallel},s_z}$), which spread over the entire ${\bm k}_{\parallel}$ region for $k_{\perp}=\pi$ and can be identified as the doping-induced states. 
In addition, Eq. (\ref{eq:overlap0}) and the validity of the decoupling approximation [Eq. (\ref{eq:overlap2})] can be confirmed using 
$$
\begin{array}{lll}
{}_0\langle{\rm GS}|{\tilde c}^{\dagger}_{-{\bm k}_{\rm F},-\varsigma}|{\rm GS}\rangle_1&=&1/\sqrt{2},\\ 
{}_0\langle{\rm GS}|S_{-{\bm k}_{\rm F}-({\bm k}_{\parallel},k_{\perp})}^{+,-\varsigma}{\tilde c}^{\dagger}_{({\bm k}_{\parallel},k_{\perp}),\varsigma}|{\rm GS}\rangle_1&=&\delta_{k_{\perp},\pi}\sqrt{2/N_{\rm s}},\\ 
{}_0\langle{\rm GS}|S_{-{\bm k}_{\rm F}-({\bm k}_{\parallel},k_{\perp})}^{z,-\varsigma}{\tilde c}^{\dagger}_{({\bm k}_{\parallel},k_{\perp}),-\varsigma}|{\rm GS}\rangle_1&=&\delta_{k_{\perp},\pi}/\sqrt{2N_{\rm s}}.
\end{array}
$$ 
\subsection{Multi-hole doping} 
A similar analysis is applicable to multi-hole-doped cases in the small-doping regime. 
The ground state at $N_{\rm h}=h$ is effectively expressed as a linear combination of direct products of $h$ $|{\rm B}_{\sigma}\rangle$ and $(N_{\rm u}-h)$ $|{\rm S}\rangle$ states. 
Because the doping-induced states are obtained by replacing one of the $h$ $|{\rm B}_{\sigma}\rangle$ states with $|{\rm T}_{s_z}\rangle$, 
the spectral weights increase as $\delta$ increases. 
In addition, because a magnetic-excitation mode similar to that of the Mott insulator remains for $\omega\approx J_{\perp}$ and because the Fermi momenta remain near $({\bm \pi},0)$, 
significant spectral weights should continue to be located almost along the momentum-shifted magnetic dispersion relation of the Mott insulator in $A_{\rm a}({\bm k}_{\parallel},\omega)$. 
The states obtained by replacing one of the $h$ $|{\rm B}_{\sigma}\rangle$ states with $|{\rm S}\rangle$ form a mode as in a doped band insulator 
($\approx$ the upper edge of the LHB at half-filling) in $A_{\rm b}({\bm k}_{\parallel},\omega)$. \cite{EderOhtaStripe} 
\par
\begin{figure*}
\includegraphics[width=17.5cm]{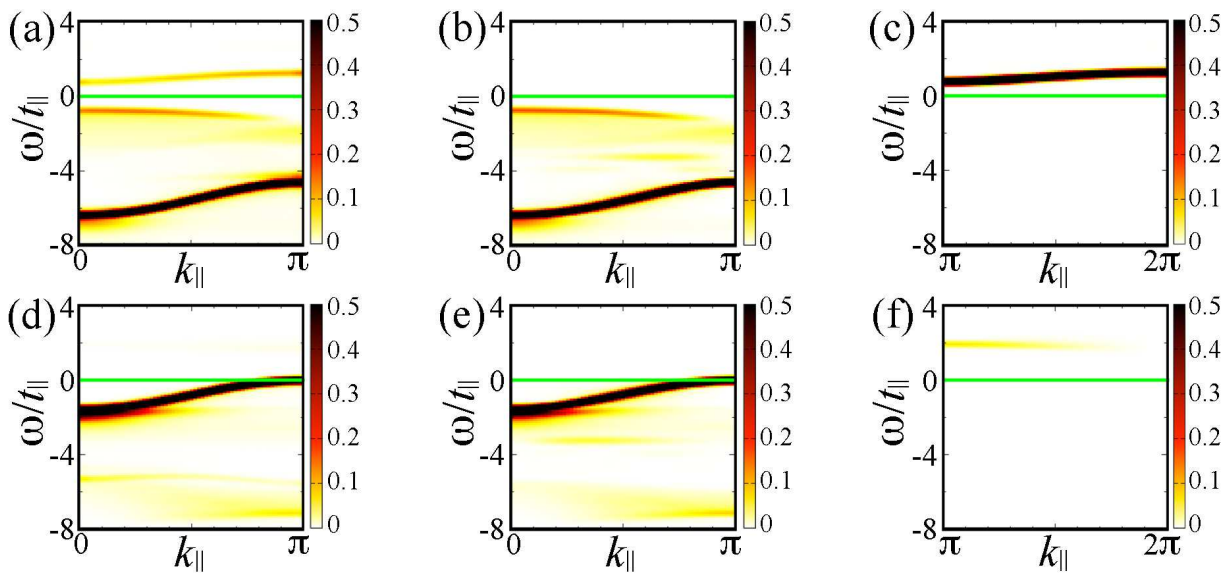}
\caption{(a) $A_{\rm a}(k_{\parallel},\omega)t_{\parallel}$ at $N_{\rm h}=2$, (b) $A_{\rm a}(k_{\parallel},\omega)t_{\parallel}$ at $N_{\rm h}=0$, and (c) $S_{\rm a}(k_{\parallel},\omega)t_{\parallel}$ at $N_{\rm h}=0$ 
in the $t$-$J$ ladder for $t_{\perp}/t_{\parallel}=2$, $J_{\parallel}/t_{\parallel}=1/4$, $J_{\perp}/t_{\parallel}=1$, and $N_{\rm s}=120$, obtained using the non-Abelian DDMRG method. 
At $N_{\rm h}=0$, $\mu$ is set to the value of the small-doping limit. 
(d)--(f) The same as (a)--(c) but for $A_{\rm b}(k_{\parallel},\omega)t_{\parallel}$ and $S_{\rm b}(k_{\parallel},\omega)t_{\parallel}$. 
The solid green lines indicate $\omega=0$. Gaussian broadening with a standard deviation of $0.1t_{\parallel}$ is used.}
\label{fig:Akw}
\end{figure*}
More simply and generally, the behavior of the spectral function at finite $\delta(=\delta_0)$ can be understood as being deformed continuously from that of the small-doping limit 
if phase transition does not occur for $0<\delta\le\delta_0$. 
This change should be small in the small-doping regime. 
\par
For confirmation, Fig. \ref{fig:Akw} shows the results for the $t$-$J$ ladder obtained using the non-Abelian dynamical density-matrix renormalization group (DDMRG) method \cite{Kohno2DtJ} 
[the DDMRG method \cite{DDMRG} in the U(1)$\otimes$SU(2) basis \cite{nonAbeliantJ,nonAbelianThesis}], where 120 density-matrix eigenstates are retained. 
The spin dynamical structure factors $S_{\rm a}(k_{\parallel},\omega)$ and $S_{\rm b}(k_{\parallel},\omega)$ are defined, respectively, in the same way as $A_{\rm a}(k_{\parallel},\omega)$ and $A_{\rm b}(k_{\parallel},\omega)$ for $\omega>0$ but for spin operators. 
The mode for $\omega/t_{\parallel}\approx 1$ in $A_{\rm a}(k_{\parallel},\omega)$ at $N_{\rm h}=2$ corresponds to the doping-induced states 
and exhibits almost the same dispersion relation as that of the magnetic excitation at $N_{\rm h}=0$, but shifted by approximately $\pi$ in $k_{\parallel}$ [Figs. \ref{fig:Akw}(a)--\ref{fig:Akw}(c)]. 
The states for $k_{\parallel}\approx\pi$ and $\omega\approx 0$ in $A_{\rm b}(k_{\parallel},\omega)$ form a mode as in a doped band insulator [Figs. \ref{fig:Akw}(d) and \ref{fig:Akw}(e)]. 
Thus, the results shown in Fig. \ref{fig:Akw} are consistent with the above analysis. 
\par
The results shown in this section (Sec.~\ref{sec:2siteClusters}) 
demonstrate that the magnetically excited states of the Mott insulator emerge as the doping-induced states in the small-doping limit in 1D and 2D systems 
and that the dispersion relation of the doping-induced states is essentially the same as that of the magnetic excitation, but shifted by the Fermi momentum. 
In the above ladder and bilayer $t$-$J$ models, the doping-induced states in the small-doping limit are disconnected from the low-energy states because the magnetic excitation at half-filling has an energy gap; 
in models having a gapless magnetic excitation at half-filling, the doping-induced states should exhibit a gapless dispersion relation in the small-doping limit. 
\section{Difference due to doping} 
The difference in the electron-addition spectra between a Mott insulator and the small-doping limit is explained as follows. 
In a Mott insulator, $c_{{\bm k},\sigma}^{\dagger}|{\rm GS}\rangle_0$ yields no spectral weights in the Mott gap 
because there are no eigenstates with $N_{\rm e}=N_{\rm s}+1$ in this energy regime. 
In the small-doping limit, in contrast, $c_{{\bm k},\sigma}^{\dagger}|{\rm GS}\rangle_1$ can yield spectral weights in the Mott gap 
because low-energy eigenstates for the magnetic excitation of the Mott insulator exist. \cite{Kohno1DHub,Kohno2DHub,KohnoSpin,Kohno2DtJ,Kohno2DHubNN,EderOhtaIPES} 
\par
The emergence of the electron-addition states in the Mott gap reflects the spin-charge separation of the Mott insulator. \cite{Kohno2DHub,KohnoSpin,Kohno2DtJ,Kohno2DHubNN} 
The spin-charge separation means that the low-energy properties are described in terms of independent spin and/or charge degrees of freedom rather than in terms of an electron-like quasiparticle. 
In 1D interacting electron systems, the spin-charge separation occurs even in a metallic phase in the low-energy limit. \cite{HaldaneTLL,TomonagaTLL,LuttingerTLL} 
On the other hand, in a Mott insulator, the spin-charge separation occurs \cite{AndersonSCsep} in any dimension: 
the low-energy properties are described in terms of the spin excitation which is gapless or has an energy gap of the order of the (effective) spin-spin interaction, 
whereas the charge excitation has an energy gap of the order of the Coulomb interaction. 
Because of this spin-charge separation, although there are low-energy magnetically excited states, 
the electron-addition excitation does not have spectral weights in the Mott gap in a Mott insulator; 
the charge quantum number of the magnetically excited states ($N_{\rm e}=N_{\rm s}$) differs from that of $c_{{\bm k},\sigma}^{\dagger}|{\rm GS}\rangle_0$ ($N_{\rm e}=N_{\rm s}+1$). 
However, the magnetically excited states can emerge in the electron-addition spectrum with the dispersion relation shifted by the Fermi momentum following the doping of the Mott insulator 
as discussed above. \cite{Kohno1DHub,Kohno2DHub,KohnoSpin,Kohno2DtJ,Kohno2DHubNN,EderOhtaIPES} 
This is in contrast with the noninteracting band insulator case, where the lowest spin excitation energy is the same as the band gap. 
\par
In the ladder and bilayer $t$-$J$ models considered above, although the ground-state behavior is similar to that of a (doped) band insulator, 
the characteristic of the Mott transition (the emergence of the states) occurs for $\omega>0$, 
reflecting the spin-charge separation of the Mott insulator (the presence of the low-energy spin excitation and the absence of the low-energy charge excitation). 
\section{Real-space picture} 
The situation in the small-doping regime can be intuitively explained as follows. 
The added electron, which is located at the position of a vacancy at low energies, has almost no chance to hop to another vacant site if vacancies are distributed homogeneously. 
However, the added electron can hop with a small energy to a neighboring site occupied by an electron having the opposite magnetization, 
at the same time as that electron hops to the site that was occupied by the added electron. 
This is nothing but the spin exchange process. 
Thus, in general, the electron-addition excitation in the small-doping regime exhibits a dispersion relation similar to that of the magnetic excitation of the Mott insulator, but shifted by the Fermi momentum. 
This shift is due to the momentum difference between the ground states before and after the electron addition. 
The spectral weight in the large-repulsion regime should be $O(\delta)$ because the probability that the added electron occupies a vacant site is essentially $N_{\rm h}/N_{\rm s}$ in the small-doping regime. 
\section{Remarks} 
Even if the ground state at half-filling has an antiferromagnetic long-range order, 
$c_{{\bm k},\sigma}^{\dagger}|{\rm GS}\rangle_1$ should have nonzero overlap with the spin-wave mode at half-filling because both involve spin-1 components if $|{\rm GS}\rangle_0$ and $|{\rm GS}\rangle_1$ have spin-0 and spin-1/2 components, respectively. 
Thus, the doping-induced states in the small-doping limit are expected to exhibit essentially the same dispersion relation 
as that of the spin-wave mode, but shifted by the Fermi momentum, in the momentum region where the LHB is not completely filled (primarily outside the Fermi surface). 
This picture is consistent with the results for the 2D Hubbard and $t$-$J$ models in Refs. \onlinecite{Kohno2DHub,KohnoSpin,Kohno2DtJ}. 
\par
The continuous transformation of the single-particle dispersion relation for $\omega>0$ into the momentum-shifted magnetic dispersion relation of the Mott insulator does not necessarily mean that 
the magnetic excitation of the Mott insulator is interpreted in terms of spinons (spin-1/2 and charge-0 quasiparticles), because this characteristic is independent of quasiparticle pictures. 
The momentum shift is simply due to the momentum difference between the ground states before and after the electron addition. 
In the above ladder and bilayer $t$-$J$ models, the magnetic excitation is typically interpreted in terms of a triplon (spin-1 and charge-0 quasiparticle). 
\par
The doping-induced states in the small-doping limit can involve not only the magnetically excited states but also other low-energy excited states (spin-0 excited states) at half-filling 
if $c_{{\bm k},\sigma}^{\dagger}|{\rm GS}\rangle_1$ contains components with these quantum numbers. 
Nevertheless, the doping-induced states are expected to be primarily due to the magnetic excitation because this is generally the most dominant low-energy excitation in a Mott insulator (Fig. \ref{fig:Akw}). \cite{Kohno1DHub,Kohno2DHub,KohnoSpin,Kohno2DtJ} 
\par
In a first-order Mott transition where phase separation occurs for $0<\delta<\delta_c(\ne 0)$, 
the states obtained by adding an electron in the $\delta\rightarrow 0$ limit are mixtures of the electron-addition states from the ground state at $\delta=\delta_c$ and those at $\delta=0$, 
which are not exactly the magnetically excited states at half-filling. 
Nevertheless, 
the dispersing mode for $\omega>0$ in the LHB before the Mott transition in the large-repulsion regime 
should more or less reflect the magnetic excitation from a Mott insulating state (an eigenstate at half-filling) similar to the ground state at $\delta=\delta_c$. 
Thus, the indications that a spin-charge-separated Mott insulator is being approached, such as a reduction in spectral weight from the dispersing mode, \cite{Kohno1DHub,Kohno2DHub,KohnoSpin,Kohno2DtJ,Kohno2DHubNN} 
are expected to generally appear near the Mott transition in the large-repulsion regime, provided the value of $\delta_c$ is relatively small. 
\section{Discussion and Summary} 
We consider the issue of the doping-induced states in the 2D Hubbard and $t$-$J$ models mentioned in Sec. \ref{sec:intro}. 
Because the low-energy magnetic excitation at half-filling is well described in terms of the gapless spin-wave mode, \cite{AndersonSW,Manousakis2dHeis,Hirsch2dHub} 
according to the present results, a mode exhibiting the spin-wave dispersion relation shifted by the Fermi momenta should emerge in the electron-addition spectrum in the small-doping limit; 
the mode should be gapless at the Fermi momenta because the spin-wave mode is gapless. 
If the LHB around $(0,0)$ is essentially completely filled, the mode appears primarily outside this momentum region. 
This picture is consistent with the results in Refs. \onlinecite{Kohno2DHub,KohnoSpin,Kohno2DtJ}; 
interpretations suggesting that the doping-induced states are essentially disconnected from the low-energy states by an energy gap 
\cite{SakaiImadaPRL,SakaiImadaPRB,SakaiImadaRaman,Tohyama2DtJ,PhillipsRMP,PhillipsMottness,EderOhtaIPES,EderOhta2DHub,EderOhtaRigidband,ImadaCofermionPRL,ImadaCofermionPRB} 
are not consistent with this picture in the small-doping limit. 
For the 1D Hubbard model, this picture essentially agrees with the results in Refs. \onlinecite{Kohno1DHub,KohnoSpin}. 
\par
In summary, the general and direct relationships of the doping-induced states with the magnetically excited states of the Mott insulator were clarified 
based on an analysis of the quantum numbers and the overlaps, as well as through a demonstration using the ladder and bilayer $t$-$J$ models. 
The results imply that the doping-induced states in the small-doping limit exhibit essentially the same dispersion relation as that of the magnetic excitation of the Mott insulator, but shifted by the Fermi momentum, 
in the momentum region where the LHB is not completely filled in a continuous Mott transition. 
This characteristic should be general and independent of the dimensionality and quasiparticle pictures. 
\par
With the reverse use of this characteristic, the magnetic dispersion relation of the Mott insulator can be inferred from the dispersion relation of the electron-addition excitation in the small-doping limit 
by shifting this dispersion relation by the Fermi momentum. 
In particular, if the electron-addition excitation has a gap in an energy regime, the magnetic excitation is also expected to have a gap in this energy regime. 
\begin{acknowledgments}
This work was supported by KAKENHI (Grants No. 23540428 and No. 26400372) and the World Premier International Research Center Initiative (WPI), MEXT, Japan. 
The numerical calculations were partly performed on the supercomputer at the National Institute for Materials Science. 
\end{acknowledgments}

\end{document}